\documentstyle[aps,epsfig]{revtex}
\begin{document}
\title{\Large\bf {Toward a global description of the nucleus-nucleus 
interaction.}}
\author{ 
L. C. Chamon$^1$, B. V. Carlson$^2$, L. R. Gasques$^1$, D. Pereira$^1$, 
C. De Conti$^2$, M. A. G. Alvarez$^1$, 
\\ M. S. Hussein$^1$, M. A. C\^andido Ribeiro$^3$, E. S. Rossi Jr.$^1$ and 
C. P. Silva$^1$.}

\address{ 
\vspace{2mm}
1. Departamento de F\'{\i}sica Nuclear, Instituto de F\'{\i}sica da 
Universidade de S\~{a}o Paulo, 
\\ Caixa Postal 66318, 05315-970, S\~{a}o Paulo, SP, Brazil.
\vspace{2mm}
\\ 2. Departamento de F\'{\i}sica, Instituto Tecnol\'{o}gico de 
Aeron\'{a}utica, Centro T\'ecnico Aeroespacial, 
\\ S\~{a}o Jos\'e dos Campos, SP, Brazil.
\vspace{2mm}
\\ 3. Departamento de F\'{\i}sica, Instituto de Bioci\^encias, Letras e 
Ci\^encias Exatas,
\\ Universidade Estadual Paulista, S\~{a}o Jos\'e do Rio Preto, SP, Brazil.}
\maketitle
\begin{abstract}
Extensive systematizations of theoretical and experimental nuclear densities 
and of optical potential strengths extracted from heavy-ion elastic scattering 
data analyses at low and intermediate energies are presented. The 
energy-dependence of the nuclear potential is accounted for within a model 
based on the nonlocal nature of the interaction. The systematics indicates 
that the heavy-ion nuclear potential can be described in a simple global way 
through a double-folding shape, which basically depends only on the density of 
nucleons of the partners in the collision. The possibility of extracting 
information about the nucleon-nucleon interaction from the heavy-ion potential 
is investigated. \\
\end{abstract}

\vspace{5mm}

{\it \noindent PACS:} 24.10.Ht, 13.75.Cs, 21.10.Ft, 21.10.Gv, 21.30.-x \\
{\it Keywords:} Heavy-ion nuclear potential. Proton, neutron, charge, nucleon 
and matter density distributions. Effective nucleon-nucleon interaction.

\vspace{5mm}

{\large \bf \noindent 1. Introduction}
\vspace{5mm} \\
\indent The optical potential plays a central role in the description of 
heavy-ion collisions, since it is widely used in studies of the elastic 
scattering process as well as in more complicated reactions through the DWBA or 
coupled-channel formalisms. This complex and energy-dependent potential is 
composed of the bare and polarization potentials, the latter 
containing the contribution arising from nonelastic couplings. In principle, 
the bare (or nuclear) potential between two heavy ions can be associated with 
the fundamental nucleon-nucleon interaction folded into a product of the 
nucleon densities of the nuclei \cite{ref0}. Apart from some structure effects, 
the shape of the nuclear density along the table of stable nuclides is nearly 
a Fermi distribution, with diffuseness approximately constant and radius 
given roughly by $R=r_0 \; A^{1/3}$, where $A$ is the number of nucleons of 
the nucleus. Therefore, one could expect a simple dependence of the heavy-ion 
nuclear potential on the number of nucleons of the partners in the collision. 
In fact, analytical formulae have been deduced \cite{ref1,ref2,ref3} for the 
folding potential, and simple expressions have been obtained at the surface 
region. An universal (system-independent) shape for the heavy-ion nuclear 
potential has been derived \cite{ref4} also in the framework of the 
liquid-drop model, from the Proximity Theorem which relates the force between 
two nuclei to the interaction between flat surfaces made of semi-infinite 
nuclear matter. The theorem leads \cite{ref4} to an expression for the 
potential in the form of a product of a geometrical factor by a function of 
the separation between the surfaces of the nuclei.

The elastic scattering is the simplest process that occurs in a heavy-ion 
collision because it involves very litle rearrangement of matter and energy. 
Therefore, this process has been studied in a large number of 
experimental investigations, and a huge body of elastic cross section data is 
currently available. The angular distribution for elastic scattering provides 
unambiguous determination of the real part of the optical potential only in a 
region around a particular distance \cite{ref5} hereafter referred as the 
sensitivity radius ($R_S$). At energies close to the Coulomb barrier the 
sensitivity radius is situated in the surface region. In this energy region, 
the systematization \cite{ref6,ref7} of experimental results for potential 
strengths at the sensitivity radii has provided an universal exponential shape 
for the heavy-ion nuclear potential at the surface, as theoretically expected, 
but with a diffuseness value smaller than that originally proposed in the 
proximity potential.

In a recent review article \cite{ref5} the phenomenon of rainbow scattering 
was discussed, and it was emphasized that the real part of the optical 
potential can be unambiguously extracted also at very short distances from 
heavy-ion elastic scattering data at intermediate energies. Such a kind of 
data has been first obtained for $\alpha$-particle scattering from a variety 
of nuclei over a large range of energies \cite{ref8,ref9,ref10}, and later for 
several heavy-ion systems. However, differently from the case for the surface 
region (low energy), a systematization of potential strengths at the inner 
distances has not been performed up to now, probably because the resulting 
phenomenological interactions have presented significant dependence on the 
bombarding energies. Several theoretical models have been developed to account 
for this energy-dependence through realistic mean field potentials. Most of 
them are improvements of the original double-folding potential with the 
nucleon-nucleon interaction assumed to be energy- and density-dependent 
\cite{ref5}. Another recent and successful model \cite{ref11,ref12,ref13} 
associates the energy-dependence of the heavy-ion bare potential with nonlocal 
quantum effects related to the exchange of nucleons between target and 
projectile, resulting in a very simple expression for the energy-dependence of 
the nuclear potential. Using the model of Refs. \cite{ref11,ref12,ref13}, in 
the present work we have realized a systematization of potential strengths 
extracted from elastic scattering data analyses, considering both: low 
(near-barrier) and intermediate energies. The systematics indicates that the 
heavy-ion nuclear potential can be described in a simple global way through a 
double-folding shape, which basically depends only on the number of nucleons 
of the nuclei.

The paper is organized as follows. In Section 2, as a preparatory step for 
the potential systematization, an extensive and systematic study of nuclear 
densities is presented. This study is based on charge distributions extracted 
from electron scattering experiments \cite{ref15,ref16} as well as on 
theoretical densities derived from the Dirac-Hartree-Bogoliubov model 
\cite{ref14}. In Section 3, analytical expressions for the double-folding 
potential are derived for the whole (surface and inner) interaction region, 
and a survey of the main characteristics of this potential is presented. 
Section 4 contains the nonlocal model for the heavy-ion bare interaction, 
including several details that have not been published before. Section 5 
is devoted to the nuclear potential systematics. In Section 6, we discuss the 
role played by the nucleon-nucleon interaction, and we present, in a somewhat 
speculative way, an alternative form for the effective nucleon-nucleon 
interaction, which is consistent with our results for the heavy-ion nuclear 
potential. Finally, Section 7 contains a brief summary and the main 
conclusions. 

\vspace{7mm} 
{\large \bf \noindent 2. Systematization of the nuclear densities}
\vspace{5mm} \\
\indent According to the double-folding model, the heavy-ion nuclear potential 
depends on the nuclear densities of the nuclei in collision. Thus, a 
systematization of the potential requires a previous systematization of the 
nuclear densities. In this work, with the aim of describing the proton, 
neutron, nucleon (proton+neutron), charge and matter densities, we adopt the 
two-parameter Fermi (2pF) distribution, which has also been commonly used for 
charge densities extracted from electron scattering experiments \cite{ref15}. 
The shape, Eq. 1 and Fig. 1, of this distribution is particularly appealing for 
the density description, due to the flatness of the inner region, that is 
associated with the saturation of the nuclear medium, and to the rapid fall-off 
(related to the diffuseness parameter $a$) that brings out the notion of the 
radius, $R_0$, of the nucleus. 
\begin{equation}
\rho(r) = \frac{\rho_0}{1+exp \left( \frac{r-R_0}{a} \right) }
\end{equation}
The $\rho_0$, $a$ and $R_0$ parameters are connected by the normalization 
condition: 
\begin{equation}
4 \pi \; \int_0^{\infty} \rho(r) \; r^2 \; dr = X \; ,
\end{equation}
where $X$ could be the number of protons $Z$, neutrons $N$ or nucleons $A=N+Z$. 
In our theoretical calculations, the charge distribution ($\rho_{ch}$) has been 
obtained by folding the proton distribution of the nucleus ($\rho_{p}$) with 
the intrinsic charge distribution of the proton in free space ($\rho_{chp}$) 
\begin{equation}
\rho_{ch}(r)=\int \; \rho_p (\vec{r'}) \; \rho_{chp} (\vec{r}-\vec{r'}) \; 
d\vec{r'} \; ,
\end{equation}
where $\rho_{chp}$ is an exponential with diffuseness $a_{chp}=0.235 \; fm$. 
In an analogous way, we have defined the matter density of the nucleus by 
folding the nucleon distribution of the nucleus with the intrinsic matter 
distribution of the nucleon, which is assumed to have the same shape of the 
intrinsic charge distribution of the proton. For convenience, the charge and 
matter distributions are normalized to the number of protons and nucleons, 
respectively. 

In order to systematize the heavy-ion nuclear densities, we have calculated 
theoretical distributions for a large number of nuclei using the 
Dirac-Hartree-Bogoliubov (DHB) model \cite{ref14}. The DHB calculations were 
performed using the NL3 parameter set \cite{nl3}. This set was obtained by 
adjusting the masses, and the charge and neutron radii of 10 nuclei in the 
region of the valley of stability, ranging from $^{16}$O to $^{214}$Pb, using 
the Dirac-Hartree-BCS (DH-BCS) model. For the cases in which they have been 
performed, calculations using this parameter set and either the DHB 
\cite{ref14} or the DH-BCS \cite{nl3,nisn,n28} model have shown very good 
agreement with experimental masses and radii. In the present work, we have also 
used the results of previous systematics for charge distributions 
\cite{ref15,ref16}, extracted from electron scattering experiments, as a 
further check of our DHB results. All the theoretical and most of the 
``experimental" densities are not exact Fermi distributions. Thus, with the 
aim of studying the equivalent diffuseness of the densities, we have 
calculated the corresponding logarithmic derivatives (Eq. 4) at the surface 
region (at $r \approx R_0+2 \; fm$).
\begin{equation}
a \approx - \frac{\rho(r)}{\frac{d\rho}{dr}}
\end{equation}
Fig. 2a shows the results for the experimental charge distributions: the 
diffuseness values spread around an average diffuseness 
$\bar{a}_c=0.53 \; fm$, with standard deviation $0.04 \; fm$. Most of this 
dispersion arises from experimental errors. Indeed, we have verified that 
different analyses (different electron scattering data set or different models 
for the charge density) for a given nucleus provide diffuseness values that 
differ from each other by about $0.03 \; fm$. Therefore, the experimental 
charge distributions are compatible, within the experimental precision, with a 
constant diffuseness value. The theoretical charge distributions present 
similar behavior (Fig. 2b), with average value slightly smaller than the 
experimental one. In this case, the observed standard deviation, $0.02 \; fm$, 
is associated with the effects of the structure of the nuclei. Despite the 
trend presented by the neutron and proton diffuseness (Fig. 2c), all the 
nucleon distributions result in very similar diffuseness values 
($\bar{a}_N=0.48 \; fm$), with standard deviation $0.025 \; fm$. Due to the 
folding procedure, the matter distributions present diffuseness values 
significantly greater ($\bar{a}_M=0.54 \; fm$) than those for the nucleon 
distributions. Taking into account that the theoretical calculations have 
slightly underestimated the experimental charge diffuseness, we consider that 
more realistic average values for the nucleon and matter density diffuseness 
are $0.50 \; fm$ and $0.56 \; fm$, respectively. A dispersion ($\sigma_a$) of 
about $0.025 \; fm$ around these average values is expected due to effects of 
the structure of the nuclei.

The root-mean-square (RMS) radius of a distribution is defined by Eq. 5: 
\begin{equation}
r_{rms}=\sqrt{\frac{\int r^2 \; \rho(r) \; d\vec{r}}{\int \rho(r) \; d\vec{r}}}.
\end{equation}
We have determined the radii $R_0$ for the 2pF distributions assuming that 
the corresponding RMS radii should be equal to those of the experimental 
(electron scattering) and theoretical (DHB) densities. The results for 
$R_0$ from theoretical charge distributions (Fig. 3b) are very similar to 
those from electron experiments (Fig. 3a). This fact indicates that the 
radii obtained through the theoretical DHB calculations are quite realistic. 
The nucleon and matter densities give very similar radii (Fig. 3d), which are 
well described by the following linear fit: 
\begin{equation}
R_0 = 1.31 \; A^{1/3} - 0.84 \; fm .
\end{equation}
Due to effects of the structure of the nuclei, the $R_0$ values spread around 
this linear fit with dispersion $\sigma_{R_0}=0.07 \; fm$, but the heavier the 
nucleus is, the smaller is the deviation. In Fig. 4 are shown the theoretical 
(DHB) nucleon densities for a few nuclei, and the corresponding 2pF 
distributions with $a=0.50 \; fm$ and $R_0$ values obtained from Eq. 6.

\vspace{7mm} 
{\large \bf \noindent 3. Essential features of the folding potential}
\vspace{5mm} \\
\indent  The double-folding potential has the form
\begin{equation}
V_{F}(R) = \int \rho_1(r_1) \; \rho_2(r_2) \; v_{NN}(\vec{R}-\vec{r_1}
+\vec{r_2}) \; d\vec{r_1} \; d\vec{r_2} \; ,
\end{equation}
where $R$ is the distance between the centers of the nuclei, $\rho_i$ are the 
respective nucleon distributions, and $v_{NN}(\vec{r})$ is the effective 
nucleon-nucleon interaction. The success of the folding model can only be 
judged meaningfully if the effective nucleon-nucleon interaction employed is 
truly realistic. The most widely used realistic interaction is known as M3Y 
\cite{ref0,ref5}, which can usually assume two versions: Reid and Paris.

For the purpose of illustrating the effects of density variations on the 
folding potential, we show in Fig. 5 the results obtained for different 
sets of 2pF distributions. In Section 2, we have estimated the dispersions of 
the $R_0$ and $a$ parameters, $\sigma_{R_0} \approx 0.07$ and 
$\sigma_a \approx 0.025 \; fm$, that arise from effects of the structure of 
the nuclei. Observe that these standard deviations are one half of the 
corresponding variations considered in the example of Fig. 5, 
$\Delta R_0 = 0.14 \; fm$ and $\Delta a = 0.05 \; fm$. The surface region 
of the potential ($R \ge R_1+R_2$) is much more sensitive to small changes of 
the density parameters than the inner region. Our calculations indicate that, 
due to such structure effects, the strength of the nuclear potential in the 
region near the barrier radius may vary by about 20\%, and the major part of 
this variation is connected to the standard deviation of the parameter $a$. 
Therefore, concerning the nuclear potential, the effects of the structure of 
the nuclei are mostly present at the surface and mainly related to the 
diffuseness parameter.

The six-dimensional integral (Eq. 7) can easily be solved by reducing it 
to a product of three one-dimensional Fourier transforms \cite{ref0}, 
but the results may only be obtained through numerical calculations. In order 
to provide analytical expressions for the folding potential, we consider, as an 
approximation, that the range of the effective nucleon-nucleon interaction is 
negligible in comparison with the diffuseness of the nuclear densities. In 
this zero-range approach, the double-folding potential can be obtained from: 
\begin{equation}
v_{NN}(\vec{r}) \approx V_0 \; \delta(\vec{r}) \Rightarrow  V_F(R)= 
\frac{2 \pi V_0}{R} \int_0^\infty r_1 \; \rho_1(r_1)  \left[ 
\int_{|R-r_1|}^{R+r_1}  r_2 \; \rho_2(r_2) \; dr_2 \right] dr_1.
\end{equation}
As discussed in Section 2, the heavy-ion densities involved in Eq. 8 are 
approximately 2pF distributions, with $R_0 \gg a$. In the limit 
$a \rightarrow 0$, the double-integral results in 
\begin{equation}
V_F(R \le R_2-R_1) = V_0 \; \rho_{01} \; \rho_{02} \; \frac{4}{3} \pi R_1^3 \; ,
\end{equation}
\begin{equation}
V_F(R_2-R_1 \le R \le R_1+R_2) = V_0 \; \rho_{01} \; \rho_{02} \; \frac{4}{3} 
\pi {\cal R}^3 \; \left( \frac{\tau^2}{1 + \zeta \tau} \right) \left[ 
\frac{3}{8} + \frac{\tau}{4} + \zeta \frac{\tau^2}{16} \right] \; , 
\end{equation}
\begin{equation}
V_F(R \ge R_1+R_2)= 0 \; , 
\end{equation}
where $s=R-(R_1+R_2$), ${\cal R}=2 R_1  R_2/(R_1+R_2)$, 
$\zeta= {\cal R}/(R_1+R_2)$, $\tau=s/{\cal R}$, $R_1$ and $R_2$ are the 
radii of the nuclei (hereafter we consider $R_2 \ge R_1$). We need a further 
approximation to obtain analytical expressions for the folding potential in 
the case of finite diffuseness value. 

The Fermi distribution may be represented, with precision better than 3\% for 
any $r$ value (see Fig. 1), by: 
\begin{equation}
\frac{\rho_0}{1+exp \left( \frac{r-R_0}{a} \right) } \approx \rho_0 \; 
C \left( \frac{r-R_0}{a} \right) \; ,
\end{equation}
\begin{equation}
C(x \le 0)= 1 - \frac{7}{8} e^x + \frac{3}{8} e^{2x} \; ,
\end{equation}
\begin{equation}
C(x \ge 0)= e^{-x} \left( 1 - \frac{7}{8} e^{-x} + \frac{3}{8} e^{-2x} \right). 
\end{equation}
This approximation is particularly useful in obtaining analytical expressions 
for integrals that involve the 2pF distribution. If both nuclei have the same 
diffuseness $a$, the double-integral (Eq. 8) can be solved analytically 
using the approximation represented by Eq. 12, and the result expressed as a 
sum of a large number of terms, most of them negligible for $a \ll R_0$. 
Rather simple expressions can be found after an elaborate algebraic 
manipulation:
\begin{equation}
V_F(R \le R_2-R_1+a) \approx V_0 \; \rho_{01} \; \rho_{02} \; \frac{4}{3} \pi 
R_1^3 \; \left\{ 1 + 9.7 \left( \frac{a}{R_1} \right)^2 - \left[ 0.875 
\left( \frac{R_2^3}{R_1^3} -1 \right) + \frac{a}{R_1} \left( 2.4 + 
\frac{R_2^2}{R_1^2} \right) \right] e^{-(R_2-R_1)/a} \right\} \; ,  
\end{equation}
\begin{eqnarray}
V_F(R_2-R_1+a \le R \le R_1+R_2) \approx V_0 \; \rho_{01} \; \rho_{02} \; & & 
\nonumber\\ 
\frac{4}{3} \pi {\cal R}^3 \; \left( \frac{1}{1+ \zeta \tau} \right) 
\left\{ \tau^2 \left[ \frac{3}{8} + \frac{\tau}{4} + \zeta \frac{\tau^2}{16} 
\right] + 2.4 \; \eta^2 \left[ 1 - \frac{5}{8} \eta - \zeta \tau^2 +  
\left( \frac{5}{4} \eta - \frac{1}{2} \right) e^{\varepsilon} + \left( 1 + 
\frac{5}{8} \eta \right) e^{-(\varepsilon+2R_1/a)} \right] \right\} \; , & &
\end{eqnarray}
\begin{equation}
V_F(R \ge R_1+R_2) \approx V_0 \; \rho_{01} \; \rho_{02} \; \pi a^2 \; {\cal R} 
\; g(\tau) \; f(s/a) \; ,
\end{equation}
\vspace{2mm}

\noindent with $\eta=a/{\cal R}$, $\varepsilon=s/a$. The functions $g$ and $f$ 
are given by:
\begin{equation}
g(\tau) = \frac{1+ \tau + \tau^2 \zeta/3 + \eta + (\eta+1/2) \; 
e^{-\varepsilon}}{1+ \zeta \tau } \; ,
\end{equation}
\begin{equation}
f(s/a) = (1 + s/a) \; e^{-s/a} \; .
\end{equation}
If the nuclei have slightly (about 10\%) different diffuseness, the formulae 
are still valid with $a \approx (a_1+a_2)/2$. As an example of the precision 
of the analytical expressions above, we exhibit in Fig. 6 the results of 
numerical calculations (Eq. 8) and compare them with those from Eqs. 15, 16 
and 17, and also with the exact expressions for $a=0$, Eqs. 9, 10 and 11.

Eq. 17 presents some similarity with the proximity potential \cite{ref4}:
\begin{equation}
V_P = 2 \pi \Gamma {\cal R} d \; \Phi (s/d) \; ,
\end{equation}
where $d$ is the ``surface width", and $\Phi$ is an universal 
(system-independent) function. For a 2pF distribution, the surface width is 
related to the diffuseness parameter through: $d \approx (\pi/\sqrt{3}) \; a$ 
\cite{ref17}. The theoretical value adopted for $d$ is $1 \; fm$ \cite{ref4}, 
which corresponds to a diffuseness $a \approx 0.55 \; fm$. Taking into account 
that the $\Gamma$ value is rather system-independent \cite{ref4}, 
systematizations of heavy-ion potential strengths extracted from elastic 
scattering data analyses have been performed by using the following expression, 
which should be valid for surface distances,
\begin{equation}
\frac{V_P(s \gg 0)}{{\cal R}} = V_0 \; e^{-s/\alpha} \; .
\end{equation}
The resulting experimental $\alpha$ values are quite similar, 
$\alpha \approx 0.62 \; fm$ \cite{ref6,ref7}, but smaller than the theoretical 
prediction of the proximity potential $\alpha \approx 0.75 \; fm$ 
\cite{ref4}. Such systematics have included only experimental potential 
strengths in the surface region, in contrast to the case of the proximity 
potential where $V/{\cal R}$ should be an universal function of $s$ also for 
inner distances. The proximity potential is not fully agreeing with our 
results for the double-folding potential in the zero-range approach (see 
Fig. 7). In fact, Eq. 17 indicates that a better choice for an universal 
quantity at the surface region would be 
\begin{equation}
V_{red}(s \ge 0) = \frac{V_F}{\rho_{01} \; \rho_{02} \; \pi a^2 {\cal{R}} \; 
g(\tau)} \; ,
\end{equation}
which results (from Eqs. 17, 19 and 22) the system-independent expression
\begin{equation}
V_{red}(s \ge 0) \approx V_0 \; (1+s/a) \; e^{-s/a} \; .
\end{equation}
However, it is not clear that one can find a simple form for such a universal 
quantity at inner distances from Eqs. 15 and 16. In Section 5, the reduced 
potential, $V_{red}$, is useful for addressig the potential strength 
systematization. Thus we define $V_{red}$ for $s \le 0$ through the following 
trivial form: 
\begin{equation}
V_{red}(s \le 0) = V_0 \; .
\end{equation}

The end of this section is devoted to the study of the effect on the folding 
potential of a finite range for the effective nucleon-nucleon interaction. The 
tri-dimensional delta function, $V_0 \; \delta(\vec{r})$, can be represented 
through the limit $\sigma \rightarrow 0$ applied to the finite-range Yukawa 
function 
\begin{equation}
Y_{\sigma}(r)= V_0 \; \frac{e^{-r/\sigma}}{4 \pi r \sigma^2} \; .
\end{equation}
Fig. 8 shows a comparison of folding potentials in the zero-range approach 
(Eq. 8) with the result obtained (from Eq. 7) using an Yukawa function 
for the effective nucleon-nucleon interaction. The finite range is not truly 
significant at small distances, and can be accurately simulated at the surface, 
within the zero-range approach, just by slightly increasing the diffuseness of 
the nuclear densities.

\vspace{7mm} 
{\large \bf \noindent 4. A nonlocal description of the nucleus-nucleus 
interaction}
\vspace{5mm} \\
\indent Before proceeding with the systematization of the potential, we first 
set the stage for the model of the heavy-ion nuclear interaction 
\cite{ref11,ref12,ref13}. When dealing with nonlocal interactions, one is 
required to solve the following integro-differential equation 
\begin{equation}
- \frac{\hbar^2}{2 \mu} \nabla^2 \Psi(\vec{R}) + \left[ V_C(R)+V_{pol}(R,E)+
\imath W_{pol}(R,E) \right] \Psi(\vec{R}) + \int U(\vec{R},\vec{R^{\prime}}) 
\; \Psi(\vec{R^{\prime}}) \; d\vec{R^{\prime}} = E \Psi(\vec{R}) \; .
\end{equation}
$V_C$ is the Coulomb interaction assumed to be local. $V_{pol}$ and $W_{pol}$ 
are the real and imaginary parts of the polarization potential, and contain 
the contribution arising from nonelastic channel couplings. The corresponding 
nonlocality, called the Feshbach nonlocality, is implicit through the 
energy-dependence of these terms, consistent with the dispersion relation 
\cite{ref20}. $U(\vec{R},\vec{R^{\prime}})$ is the bare interaction and the 
nonlocality here, the Pauli nonlocality, is solely due to the Pauli exclusion 
principle and involves the exchange of nucleons between target and projectile.

Guided by the microscopic treatment of the nucleon-nucleus scattering 
\cite{ref19,ref21,ref22,ref23,ref24}, the following ansatz is assumed for the 
heavy-ion bare interaction \cite{ref12} 
\begin{equation}
U(\vec{R},\vec{R^{\prime}}) = V_{NL} \left( \frac{R+R^{\prime}}{2} \right) 
\frac{1}{\pi^{3/2}b^3} e^{ - \left(|\vec{R}+\vec{R^{\prime}}|/b \right)^2} \; ,
\end{equation}
where $b$ is the range of the Pauli nonlocality. Introduced in this way, the 
nonlocality is a correction to the local model, and in the $b \rightarrow 0$ 
limit Eq. 26 reduces to the usual Schr\"oedinger differential equation. The 
range of the nonlocality can be found through $b \approx b_0 m_0/\mu$ 
\cite{ref18}, where $b_0 = 0.85 \; fm$ is the nucleon-nucleus nonlocality 
parameter \cite{ref19}, $m_0$ is the nucleon mass, and $\mu$ is the reduced 
mass of the nucleus-nucleus system. This type of very mild nonlocality in the 
nucleon-nucleus and nucleus-nucleus interaction is to be contrasted with the 
very strong nonlocality found in the pion-nucleus interaction in the 
$\Delta$-region \cite{ref99}. In such cases, even the concept of an optical 
potential becomes dubious. In our case, however, we are on very safe ground. 

The relation between the nonlocal interaction and the folding potential is 
obtained from \cite{ref12}
\begin{equation}
V_{NL}(R) = V_F(R) \; .
\end{equation}
Due to the central nature of the interaction, it is convenient to write down 
the usual expansion in partial waves, 
\begin{equation}
\Psi(\vec{R}) = \sum \imath^{\ell} (2 \ell +1) \frac{u_{\ell}(R)}{kR} \; 
P_{\ell}[cos(\theta)] \; ,
\end{equation}
\begin{equation}
U(\vec{R},\vec{R^{\prime}}) = \sum \frac{2 \ell +1}{4 \pi R R^{\prime}} 
\; V_{\ell}(R,R^{\prime}) \; P_{\ell}[cos(\phi)] \; ,
\end{equation}
\begin{equation}
V_{\ell}(R,R^{\prime}) = V_{NL} \left( \frac{R+R^{\prime}}{2} \right) 
\frac{1}{b \pi^{1/2}} \left[ Q_{\ell} \left( \frac{2RR^{\prime}}{b^2} \right)
e^{-\left( \frac{R-R^{\prime}}{b} \right)^2} (-)^{\ell+1} Q_{\ell} \left( 
\frac{-2RR^{\prime}}{b^2} \right)e^{-\left( \frac{R+R^{\prime}}{b} \right)^2} 
\right] \; , 
\end{equation}
where $Q_{\ell}$ are polynomials and $\phi$ is the angle between 
$\vec{R}$ and $\vec{R^{\prime}}$ \cite{ref19} . Thus, the integro-differential 
equation can be recast into the following form
\begin{equation}
\frac{\hbar^2}{2 \mu} \frac{d^2 u_{\ell}(R)}{dR^2} + 
\left[ E-V_C(R) - V_{pol}(R,E) - \imath W_{pol}(R,E) - \frac{\ell (\ell+1) 
\hbar^2}{2 \mu R^2} \right] u_{\ell}(R) = \int_0^{\infty} V_{\ell}(R, 
R^{\prime}) \; u_{\ell}(R^{\prime}) \; dR^{\prime} \; .
\end{equation}

When confronting theory and experiment, one usually relies on the optical model 
with a local potential. This brings into light the issue of extracting from 
Eq. 32 a local-equivalent (LE) potential 
\begin{equation}
V_{LE}(R,E) + \imath W_{LE}(R,E) = \frac{1}{u_{\ell}(R)} \; \int_0^{\infty} 
V_{\ell}(R, R^{\prime}) \; u_{\ell}(R^{\prime}) \; dR^{\prime} \; .
\end{equation}
The presence of the wave-function in Eq. 33 indicates that the LE potential is 
complex and also $\ell$- and energy-dependent. Despite its complex nature, the 
LE potential is not absorptive, $\langle \Psi | W_{LE} | \Psi \rangle = 0$; 
this statement can be demonstrated by considering that the nonlocal 
interaction is real and symmetrical, 
$V_{\ell}(R,R^{\prime})=V_{\ell}(R^{\prime},R)$. For neutron-nucleus systems, 
the LE potential is only weakly $\ell$-dependent, and an approximate relation 
to describe its energy-dependence has been obtained \cite{ref19}. A 
generalization of this relation for the ion-ion case is given by 
\cite{ref11,ref12}:
\begin{equation}
V_{LE}(R,E) \approx V_F(R) \; e^{- \gamma \left[ E - V_C(R) - V_{LE} (R,E) 
\right] } \; ,
\end{equation}
with $\gamma = \mu b^2/2 \hbar^2$. In order to provide an example 
of the precision of expression 34, in Fig. 9 the corresponding result is 
compared to the exact LE potential (Eq. 33) obtained from the numerical 
resolution \cite{ref12} of the respective integro-differential equations 
(Eq. 32). The local-equivalent potential is quite well described by 
Eq. 34 for any $\ell$ value, except at very small distances ($R \approx 0$) 
that are not probed by heavy-ion experiments.

Expression 34 has accounted for the energy-dependence of experimentally 
extracted potential strengths for several systems in a very large energy range 
\cite{ref11,ref12,ref13}. At near-barrier energies, 
$E \approx V_C(R_B)+V_{LE}(R_B)$, the effect of the Pauli nonlocality is 
negligible and $V_{LE}(R,E) \approx V_F(R)$, but the higher the energy is, the 
greater is the effect. At energies about $200 \; MeV/nucleon$ the 
local-equivalent potential is about 1 order of magnitude less intense than the 
corresponding folding potential (see examples in Refs. \cite{ref11,ref12}). 
In a classical physics framework, the exponent in Eq. 34 is related to the 
kinetic energy ($E_k$) and to the local relative speed between the nuclei 
($v$) by
\begin{equation}
v^2 = \frac{2}{\mu}E_k(R) = \frac{2}{\mu} \left[ E-V_C(R)-V_{LE}(R,E) \right] 
\; ;
\end{equation}
and Eq. 34 may be rewritten in the following form
\begin{equation}
V_{LE}(R,E) \approx V_F(R) \; e^{-\left[ m_0 b_0 v/(2 \hbar) \right]^2} 
\approx V_F(R) \; e^{-4 v^2/c^2} \; ,
\end{equation}
where $c$ is the speed of light. Therefore, in this context the effect of the 
Pauli nonlocality is equivalent to a velocity-dependent nuclear interaction 
(Eq. 36). Another possible interpretation is that the local-equivalent 
potential may be associated directly with the folding potential (Eq. 37), with 
an effective nucleon-nucleon interaction (Eq. 38) dependent on the relative 
speed ($v$) between the nucleons 
\begin{equation}
V_{LE}(R,E) = V_{F} = \int \rho_1(r_1) \; \rho_2(r_2) \; v_{NN}(v, 
\vec{R}-\vec{r_1}+\vec{r_2}) \; d\vec{r_1} \; d\vec{r_2} \; ,
\end{equation}
\begin{equation}
v_{NN} (v, \vec{r}) = v_{f}(\vec{r}) \; e^{-4v^2/c^2} \; .
\end{equation}

\vspace{7mm} 
{\large \bf \noindent 5. The systematization of the nuclear potential}
\vspace{5mm} \\
\indent As already mentioned, the angular distribution for elastic scattering 
provides an unambiguous determination of the real part of the optical potential 
in a region around the sensitivity radius ($R_S$). For bombarding energies 
above (and near) the barrier, the sensitivity radius is rather 
energy-independent and close to the barrier radius ($R_B$), while at 
intermediate energies much inner distances are probed. At sub-barrier energies, 
the $R_S$ is strongly energy-dependent, with its variation connected to the 
classical turning point; this fact has allowed the determination of the 
potential in a wide range of near-barrier distances, 
$R_B \le R_S \le R_B + 2 \; fm$. With the aim of avoiding ambiguities in the 
potential systematization, we have selected ``experimental" (extracted from 
elastic scattering data analyses) potential strengths at the corresponding 
sensitivity radii, from works in which the $R_S$ has been determined or at 
least estimated. In several articles, the authors claim that their data 
analyses 
at intermediate energies have unambiguously determined the nuclear potential 
in a quite extensive region of interaction distances. In such cases, we have 
considered potential strength ``data" in steps of $1 \; fm$ over the whole 
probed region. Tables 1 and 2 provide the systems included in the nuclear 
potential systematics for the sub-barrier and intermediate energies, 
respectively. For the energy region above (and near) the barrier, the 
present systematics contains potential strengths for a large number of 
different heavy-ion systems from the previous Christensen and Winther's 
systematization \cite{ref6}. Our systematics is not even near to being 
complete, but it is rather extensive and diversified enough to account well 
for the very large number of data that have been obtained in the last decades.

The experimental potential strengths represent the real part of the optical 
potential, which corresponds to the addition of the bare and polarization 
potentials. The contribution of the polarization to the optical potential 
depends on the particular features of the reaction channels involved in the 
collision, and is therefore quite system-dependent. If this contribution 
were very significant, it would be too difficult for one to set a global 
description of the heavy-ion nuclear interaction. In the present work, we 
neglect the real part of the polarization potential and associate the 
experimental potential strengths ($V_{Exp}$) with the bare interaction 
($V_{LE}$). The success of our findings seems to support such a hypothesis. 

In analysing experimental potential results for such a wide energy range and 
large number of different systems, we consider quite appropriate the use 
of system- and energy-independent quantities. We have removed the 
energy-dependence from the experimental potential strengths through the 
calculation of the corresponding folding potential strengths, $V_{F-Exp}$, 
based on Eq. 34. The system-dependence of the potential data set has then been  
removed with the use of the experimental reduced potential, $V_{red-Exp}$. For 
$s \ge 0$ this quantity was calculated from Eq. 22, and for inner $s$ values 
we have adopted the following simple definition 
\begin{equation}
V_{red-Exp}= V_0 \; \frac{V_{F-Exp}}{V_{F-Teo}} \; ,
\end{equation}
with $V_{F-Teo}$ calculated through Eq. 8. The other useful quantity is the 
distance between surfaces: $s=R_S-(R_1+R_2)$, where $R_S$ is associated to the 
sensitivity radius, and the radii of the nuclei have been obtained from Eq. 6. 

In Fig. 10 (top), the experimental reduced potential strengths are confronted 
with the theoretical prediction (Eqs. 23 and 24) for different diffuseness 
values. The fit to the data in the inner region ($s \le 0$) results 
unambiguously in the value $V_0 = - 456 \; MeV \; fm^3$, and is quite 
insensitive to the diffuseness parameter, in agreement with the discussion 
about the folding features of Section 3. The fit for $s \ge 0$ is sensitive to 
both: $V_0$ and $a$, and the corresponding best fit values are 
$a = 0.56 \; fm$ and the same $V_0$ found for the inner region. The standard 
deviation of the data set around the best fit (solid line in Fig. 10 - top) is 
25\%, a value somewhat greater than the dispersion (20\%) expected to arise 
from effects of the structure of the nuclei (as discussed in Section 3). We 
believe that the remaining difference comes from two sources: uncertainties of 
the experimentally extracted potential strengths and the contribution of the 
polarization potential that we have neglected in our analysis. We point out 
that the best fit diffuseness value, $a=0.56 \; fm$, is equal to the average 
diffuseness found (Section 2) for the matter distributions and greater than 
the average value ($a=0.50 \; fm$) of the nucleon distributions. This is a 
consistent result because we have calculated the reduced potential strengths 
based on the zero-range approach (through Eqs. 8, 22, 23 and 24). As discussed 
in Section 3, the effect of a finite-range for the effective nucleon-nucleon 
interaction can be simulated, within the zero-range approach, by increasing the 
diffuseness of the (nucleon) densities of the nuclei. This subject is dealt 
with more deeply in the next Section.

In order to characterize the importance of the Pauli nonlocality, in Fig. 10 
(bottom) are shown the results for the reduced potential through calculations 
performed without the correction (Eq. 34) due to the energy-dependence of the 
LE potential, i.e. associating the experimental potential strengths directly 
with the folding potential. The quality of the corresponding fit (Fig. 10 - 
bottom) is similar to that obtained with the nonlocality (Fig. 10 - top), but 
the $V_0$ and $a$ parameters are significantly different. In the next Section, 
we show that the values found without considering the nonlocality, 
$a=0.61 \; fm$ and $V_0= - 274 \; MeV \; fm^3$, seem to result in an 
unrealistic nucleon-nucleon interaction.

\vspace{7mm} 
{\large \bf \noindent 6. The effective nucleon-nucleon interaction}
\vspace{5mm} \\
\indent After removing the energy-dependence of the experimental 
potential strengths, the corresponding results are compatible with the 
double-folding potential in the zero-range approach (Eq. 8), provided that 
the matter densities of the nuclei be adopted in the folding procedure instead 
of the nucleon densities. In this section, we study the consistency of our 
results for the nuclear potential in the case that the double-folding model is 
treated in the more common interpretation: the nucleon distributions and a 
finite-range nucleon-nucleon interaction are assumed in Eq. 7. With the 
purpose of keeping the comparison between experimental and theoretical results 
through the use of system-independent quantities, it is necessary to change 
the definition of the experimental reduced potential
\begin{equation}
V_{red-Exp}= V_{red-Teo} \; \frac{V_{F-Exp}}{V_{F-Teo}} \; ,
\end{equation}
where $V_{F-Teo}$ is now calculated through Eq. 7. $V_{red-Teo}$ is still 
obtained from Eqs. 23 and 24, with the $V_0$ parameter being associated to the 
volume integral of the effective nucleon-nucleon interaction (actually, this 
same procedure has also been adopted in the zero-range case) 
\begin{equation}
V_0 = 4 \pi \int v_{NN}(r) \; r^2 \; dr \; .
\end{equation}

The effective nucleon-nucleon interaction should be based upon a realistic 
nucleon-nucleon force, since our goal is to obtain a unified description of 
the nucleon-nucleon, nucleon-nucleus and nucleus-nucleus scattering 
(a discussion about the ``realism" of the interaction is found in Refs. 
\cite{ref0,ref5}). For instance, a realistic interaction should match the 
empirical values for the volume integral and root-mean-square radius of the 
nucleon-nucleon interaction, $V_0 \approx -430 \; MeV \; fm^3$ and 
$r_{rms} \approx 1.5 \; fm$, that were extrapolated from the main features of 
the optical potential for the nucleon-nucleus scattering at 
$E_{nucleon} = 10 \; MeV$ \cite{ref0,ref42,ref43,ref44}. The M3Y interaction 
has been derived \cite{ref0} with basis on the {\it G}-matrix for two nucleons 
bound near the Fermi surface, and certainly is representative of realistic 
interactions. In table 3 are presented the volume integral and root-mean-square 
radius for several nucleon-nucleon interactions used in this work, including 
the M3Y at $10 \; MeV/nucleon$.

The M3Y interaction is not truly appropriate for use in the context of the 
nonlocal model, because it already contains a simulation of the exchange 
effects included in its knock-on term. Furthermore, according to the nonlocal 
model the energy-dependence of the local-equivalent potential should be 
related only to the finite range of the Pauli nonlocality, but the knock-on 
exchange term in the M3Y interaction is also energy-dependent. Therefore, the 
use of the M3Y in the nonlocal model would imply a double counting of the 
energy-dependence that arises from exchange effects. In Section 4, we have 
demonstrated that the LE potential is identical with the double-folding 
potential for energies near the barrier, which are in a region around 10 
$MeV/nucleon$. In this same energy range, the folding potential with the M3Y 
interaction have provided a very good description of elastic scattering data 
for several heavy-ion systems \cite{ref0}. Thus, we believe that an 
appropriate nucleon-nucleon interaction for the nonlocal model could be the 
M3Y ``frozen" at $10 \; MeV/nucleon$ \cite{ref12}, i.e. considering the 
parameters of the Reid and Paris versions as energy-independent values. Fig. 
11 (top) shows a comparison between experimental and theoretical heavy-ion 
reduced potentials, in which the ``frozen" M3Y-Reid was considered for the 
nucleon-nucleon interaction. We emphasize that no adjustable parameter has 
been used in these calculations, but even so a good agreement between data and 
theoretical prediction has been obtained. The ``frozen" M3Y-Paris provides 
similar results.

With the aim of investigating how much information about the effective 
nucleon-nucleon interaction can be extracted from our heavy-ion potential 
systematics, we have considered other possible functional forms for this 
effective interaction. Besides the Yukawa function (Eq. 25), we have also used 
the Gaussian (Eq. 42) and the exponential (Eq. 43), which reduce to 
the tri-dimensional delta function in the limit $\sigma \rightarrow 0$, 
\begin{equation}
G_{\sigma}(r)=V_0 \; \frac{e^{-r^2/2\sigma^2}}{(2 \pi)^{3/2} \; \sigma^3} \; ,
\end{equation}
\begin{equation}
E_{\sigma}(r)=V_0 \; \frac{e^{-r/\sigma}}{8 \pi \sigma^3} \; .
\end{equation}
The fits obtained with all these functions are of similar quality and 
comparable with that for the M3Y interaction (Fig. 11 - top). The resulting 
best fit widths ($\sigma$), volume integrals and corresponding 
root-mean-square radii are found in table 3. All the $V_0$ and $r_{rms}$ 
values, including those of the M3Y, are quite similar. Also the 
``experimentally" extracted intensity of the nucleon-nucleon interaction in 
the region $1 \le r \le 3 \; fm$ seems to be rather independent of the model 
assumed for this interaction (see Fig. 12).

In Section 5, we have demonstrated that the major part of the ``finite-range" 
of the heavy-ion nuclear potential is related only to the spatial extent of 
the nuclei. In fact, even considering a zero-range for the interaction 
$v_{NN}$ in Eq. 8, the shape of the heavy-ion potential could be well 
described just by folding the matter densities of the two nuclei. One would 
ask whether the finite-range shape of the effective nucleon-nucleon 
interaction can be derived in a similar way. Thus, we have considered a 
folding-type effective nucleon-nucleon interaction built from: 
\begin{equation}
v_{NN}(\vec{r}) \approx v_f(r)= \int \rho_m(r_1) \; \rho_m(r_2) \; 
V_0 \;\delta(\vec{R}-\vec{r_1}+\vec{r_2}) \; d\vec{r_1} \; d\vec{r_2}= \; 
\frac{2 \pi V_0}{r} \int_0^\infty r_1 \; \rho_m(r_1)  \left[ 
\int_{|r-r_1|}^{r+r_1}  r_2 \; \rho_m(r_2) \; dr_2 \right] dr_1 \; ,
\end{equation}
where $V_0 = - 456 \; MeV \; fm^3$ as determined by the heavy-ion potential 
analysis, and $\rho_m$ is the matter density of the nucleon. Based on the 
intrinsic charge distribution of the proton in free space, which has been 
determined by electron scattering experiments, we have assumed an exponential 
shape for the matter density of the nucleon
\begin{equation}
\rho_m(r) = \rho_0 \; e^{-r/a_m} \; .
\end{equation}
Of course, $\rho_0$ and $a_m$ are connected by the normalization condition, 
Eq. 2. The integration of Eq. 44 results in
\begin{equation}
v_f(r) = V_0 \; \pi \; a_m^3 \; \rho_0^2 \; e^{-r/a_m} \; 
\left( 1 + \frac{r}{a_m} + \frac{r^2}{3a_m^2} \right) \; . 
\end{equation}
With this finite-range folding-type effective nucleon-nucleon interaction, 
a good fit of the reduced heavy-ion potential strengths is obtained 
(see Fig. 11 - bottom), with realistic volume integral and root-mean-square 
radius (see table 3). The folding-type interaction is quite similar to both 
versions of the M3Y interaction in the surface region (see Fig. 12). 

The folding-type interaction in the context of the nonlocal model provides a 
very interesting unification between the descriptions of the nucleus-nucleus, 
nucleon-nucleus and effective nucleon-nucleon interactions. This can be 
appreciated through the comparison between Eqs. 36 and 38, with the subtle 
detail that  $V_F$ (in Eq. 36) and $v_f$ (in Eq. 38) can both be calculated by 
folding the matter densities in the zero-range approach, and with the same 
$V_0$ value. Therefore, the interaction between two nuclei (or nucleons) can 
be obtained from 
\begin{equation}
V_{LE}(R) = \int \rho_1(r_1) \; \rho_2(r_2) \; 
V_0 \;\delta(\vec{R}-\vec{r_1}+\vec{r_2}) \; e^{-4v^2/c^2} \; 
d\vec{r_1} \; d\vec{r_2}
\end{equation}
where $V_0=-456 \; MeV \; fm^3$, $\rho_i$ are the matter densities, and $v$ is 
the relative speed between the nuclei (or nucleons). An alternative way to 
calculate the heavy-ion interaction is with the Eq. 37 (and 38), but in this 
case the nucleon distributions must be used (in Eq. 37) instead of the matter 
densities. All these findings seems to be quite consistent. However, the best 
fit value obtained for the diffuseness ($a_m=0.30 \; fm$) of the matter 
density of the nucleon inside the nucleus is considerable greater than that 
($a_{chp}=0.235 \; fm$) found for the charge distribution of the proton in 
free space. This finding is consistent with the swelling of the nucleon 
observed in the EMC effect \cite{ref97}, but should be contrasted with the 
opposite picture of a smaller nucleon inside the nucleus as advanced within 
the concept of color transparency \cite{ref98}.

Finally, we mention that, if the energy-dependence of the Pauli nonlocality is 
not taken into account and the experimental potential strengths are associated 
directly with the folding potential, our calculations indicate that the 
corresponding effective nucleon-nucleon interaction should have the following 
unrealistic values: $V_0 \approx -270 \; MeV \; fm^3$ and 
$r_{rms} \approx 1.9 \; fm$.

\vspace{7mm} 
{\large \bf \noindent 7. Conclusion}
\vspace{5mm} \\
\indent The experimental potential strengths considered in the present 
systematics have been obtained at the corresponding sensitivity radii, a region 
where the nuclear potential is determined from the data analyses with the 
smallest degree of ambiguity. The Fermi distribution was assumed to represent 
the nuclear densities, with parameters consistent with an extensive amount of 
theoretical (DHB calculations) and experimental (electron scattering 
experiments) results. The potential data set is well described in the context 
of the nonlocal model, by the double-folding potential in the zero-range as 
well as in the finite-range approaches. The dispersion of the potential data 
around the theoretical prediction is 25\%, which is compatible with the 
expected effects arising from the variation of the densities due to the 
structure of the nuclei. If the nonlocal interaction is assumed, the heavy-ion 
potential data set seems to determine a few characteristics of the effective 
nucleon-nucleon interaction, such as volume integral and root-mean-square 
radius, in a model-independent way. 

The description of the bare potential presented in this work is based only on 
two fundamental ideas: the folding model and the Pauli nonlocality. We have 
avoided as much as possible the use of adjustable parameters, and in the case 
of the ``frozen" M3Y interaction no adjustable parameters were necessary 
to fit the experimental potential strengths. Nowadays, the other important part 
of the heavy-ion interaction, the polarization potential, is commonly treated 
within a phenomenological approach, with several adjustable parameters which 
usually are energy-dependent and vary significantly from system to system. The 
association of the nonlocal bare potential presented in this work with a more 
fundamental treatment of the polarization should be the next step toward a 
global description of the nucleus-nucleus interaction. 

\vspace{5mm}
This work was partially supported by Financiadora de Estudos e Projetos (FINEP),
Funda\c{c}\~{a}o de Amparo \`a Pesquisa do Estado de S\~ao Paulo (FAPESP), 
and Conselho Nacional de Desenvolvimento Cient\'{\i}fico e Tecnol\'ogico 
(CNPq).

\newpage

\indent {\bf Table 1:} The table presents the systems, sub-barrier bombarding 
energies, and corresponding references, that have been included in the nuclear 
potential systematics. 
\begin{center}
\begin{tabular}{|c|c|c|} \hline
System&$E_{Lab}$ ($MeV$)&Reference\\ \hline
$^{16}$O + $^{58}$Ni&35, 35.5, 36, 36.5, 37, 38& \cite{ref7,ref26}\\ 
\hline
$^{16}$O + $^{60}$Ni&35, 35.5, 36, 37, 38& \cite{ref25,ref26}\\ \hline
$^{16}$O + $^{62,64}$Ni&34, 35, 36& \cite{ref26}\\ \hline
$^{16}$O + $^{88}$Sr&43, 44, 45& \cite{ref27}\\ \hline
$^{16}$O + $^{90}$Zr&46, 47, 48& \cite{ref27}\\ \hline
$^{16}$O + $^{92}$Zr&45, 46, 47, 48& \cite{ref7,ref27}\\ \hline
$^{16}$O + $^{92}$Mo&48, 48.5, 49& \cite{ref27}\\ \hline
$^{16}$O + $^{120}$Sn&53, 54, 55& \cite{ref7}\\ \hline
$^{16}$O + $^{138}$Ba&54, 55, 56, 57& \cite{ref7}\\ \hline
$^{16}$O + $^{208}$Pb&74, 75, 76, 77, 78& \cite{ref7}\\ \hline
$^{18}$O + $^{58}$Ni&35.1, 35.5, 37.1, 38& \cite{ref29}\\ \hline
$^{18}$O + $^{60}$Ni&34.5, 35.5, 37.1, 38& \cite{ref29}\\ \hline
\end{tabular}
\end{center}

\vspace{7mm}

\indent {\bf Table 2:} The same of table 1, but for intermediate energies.
\begin{center}
\begin{tabular}{|c|c|c|} \hline
System&$E_{Lab}$ ($MeV$)&Reference\\ \hline
$p$ + $^{40}$Ca, $^{208}$Pb&30.3& \cite{ref30}\\ \hline
$d$ + $^{40}$Ca, $^{208}$Pb&52& \cite{ref30}\\ \hline
$^{4}$He + $^{40}$Ca, $^{208}$Pb&104& \cite{ref30}\\ \hline
$^{6}$Li + $^{12}$C, $^{28}$Si&210, 318& \cite{ref31,ref32}\\ \hline
$^{6}$Li + $^{40}$Ca, $^{58}$Ni, $^{90}$Zr, $^{208}$Pb&210& \cite{ref33}\\ 
\hline
$^{7}$Li + $^{12}$C, $^{28}$Si&350& \cite{ref34}\\ \hline
$^{12}$C + $^{12}$C&300, 360, 1016, 1440, 2400& \cite{ref35,ref36,ref37}\\ 
\hline
$^{12}$C + $^{208}$Pb&1440& \cite{ref37}\\ \hline
$^{13}$C + $^{208}$Pb&390& \cite{ref36}\\ \hline
$^{16}$O + $^{16}$O&250, 350, 480, 704, 1120& \cite{ref39,ref40}\\ \hline
$^{16}$O + $^{12}$C, $^{28}$Si, $^{40}$Ca, $^{90}$Zr, $^{208}$Pb&1504& 
\cite{ref38}\\ \hline
$^{40}$Ar + $^{60}$Ni, $^{120}$Sn, $^{208}$Pb&1760& \cite{ref41}\\ \hline
\end{tabular}
\end{center}

\vspace{7mm}

\indent {\bf Table 3:} The width, volume integral and root-mean-square radius 
for several effective nucleon-nucleon interactions considered in this work.
\begin{center}
\begin{tabular}{|c|c|c|c|} \hline
Interaction&$\sigma$ or $a_m$ ($fm$)&$V_0$ ($MeV \; fm^3)$&$r_{rms}$ ($fm$)\\ 
\hline
M3Y-Reid& - &- 408&1.62\\ \hline
M3Y-Paris& - &- 447&1.60\\ \hline
Yukawa&0.58&- 439&1.42\\ \hline
Gaussian&0.90&- 448&1.56\\ \hline
Exponential&0.43&- 443&1.49\\ \hline
Folding-type&0.30&- 456&1.47\\ \hline
\end{tabular}
\end{center}

\newpage

\begin{figure} 
\vspace{0.0cm}
\hspace{2.0cm}
\includegraphics{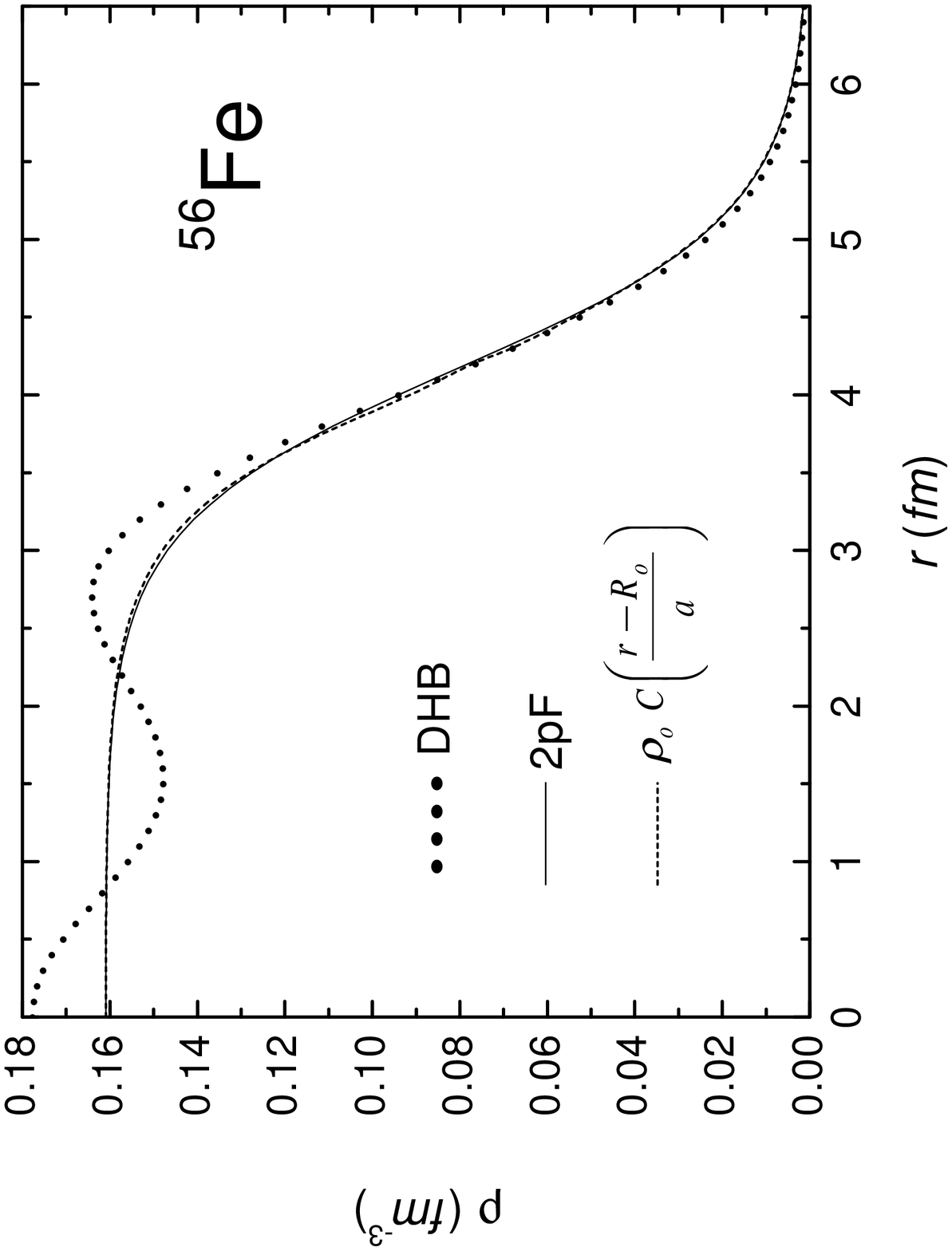}
\vspace{11.0cm}
\noindent
\caption{Nucleon density for the $^{56}$Fe nucleus represented through 
Dirac-Hartree-Bogoliubov calculations (DHB) and a two-parameter Fermi 
distribution (2pF), with $a=0.5 \; fm$ and $R_0=4.17 \; fm$. The small 
difference between the 2pF distribution and the function 
$\rho_0 \; C \left( \frac{r-R_0}{a} \right)$ (Eqs. 12, 13 and 14) is hardly 
seen in the figure.}
\end{figure}

\newpage

\begin{figure} 
\vspace{0.0cm}
\hspace{2.0cm}
\includegraphics{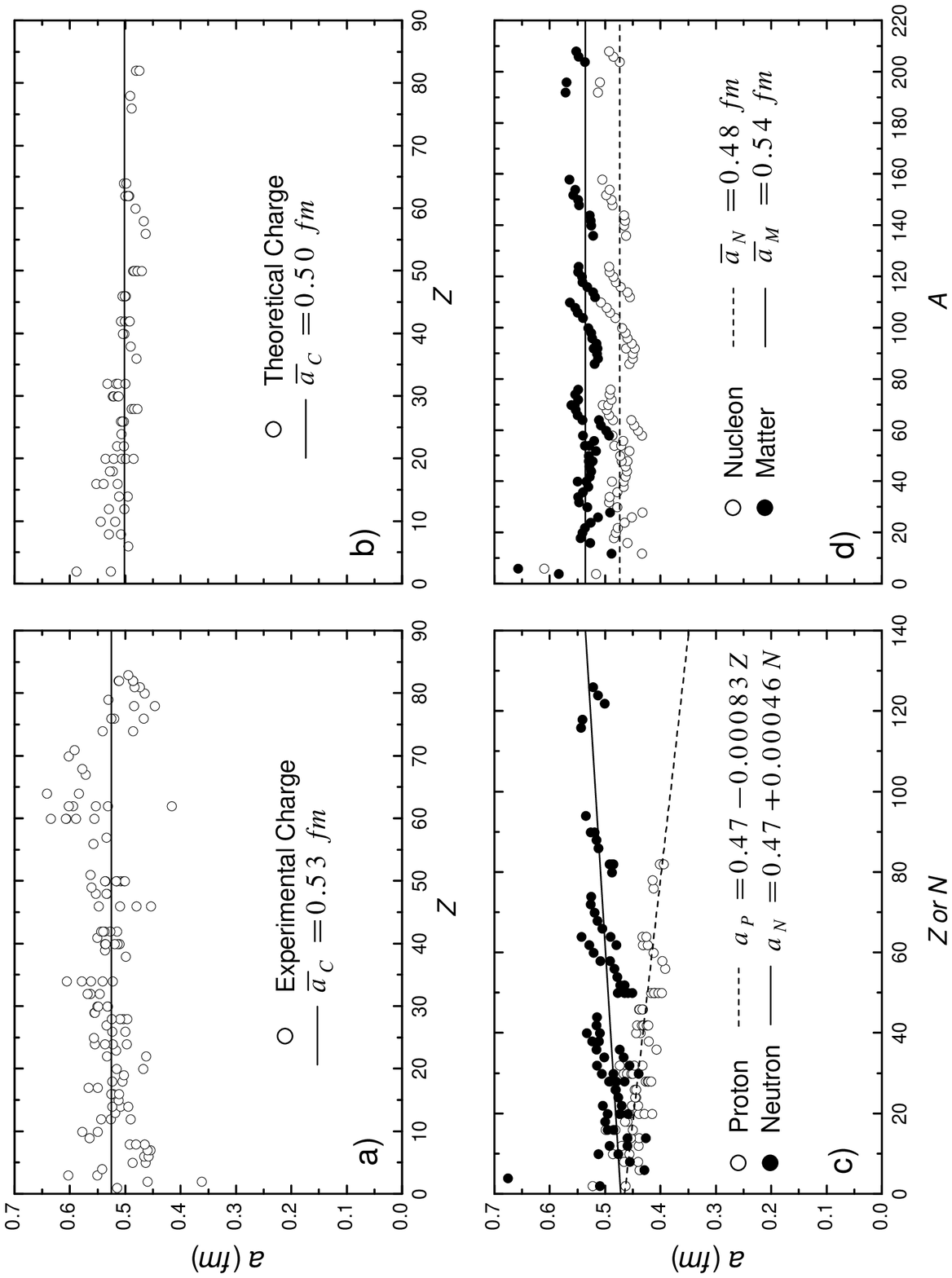}
\vspace{11.0cm}
\noindent
\caption{Equivalent diffuseness values obtained for charge distributions 
extracted from electron scattering experiments and for theoretical densities 
obtained from Dirac-Hartree-Bogoliubov calculations.}
\end{figure}

\newpage

\begin{figure} 
\vspace{0.0cm}
\hspace{2.0cm}
\includegraphics{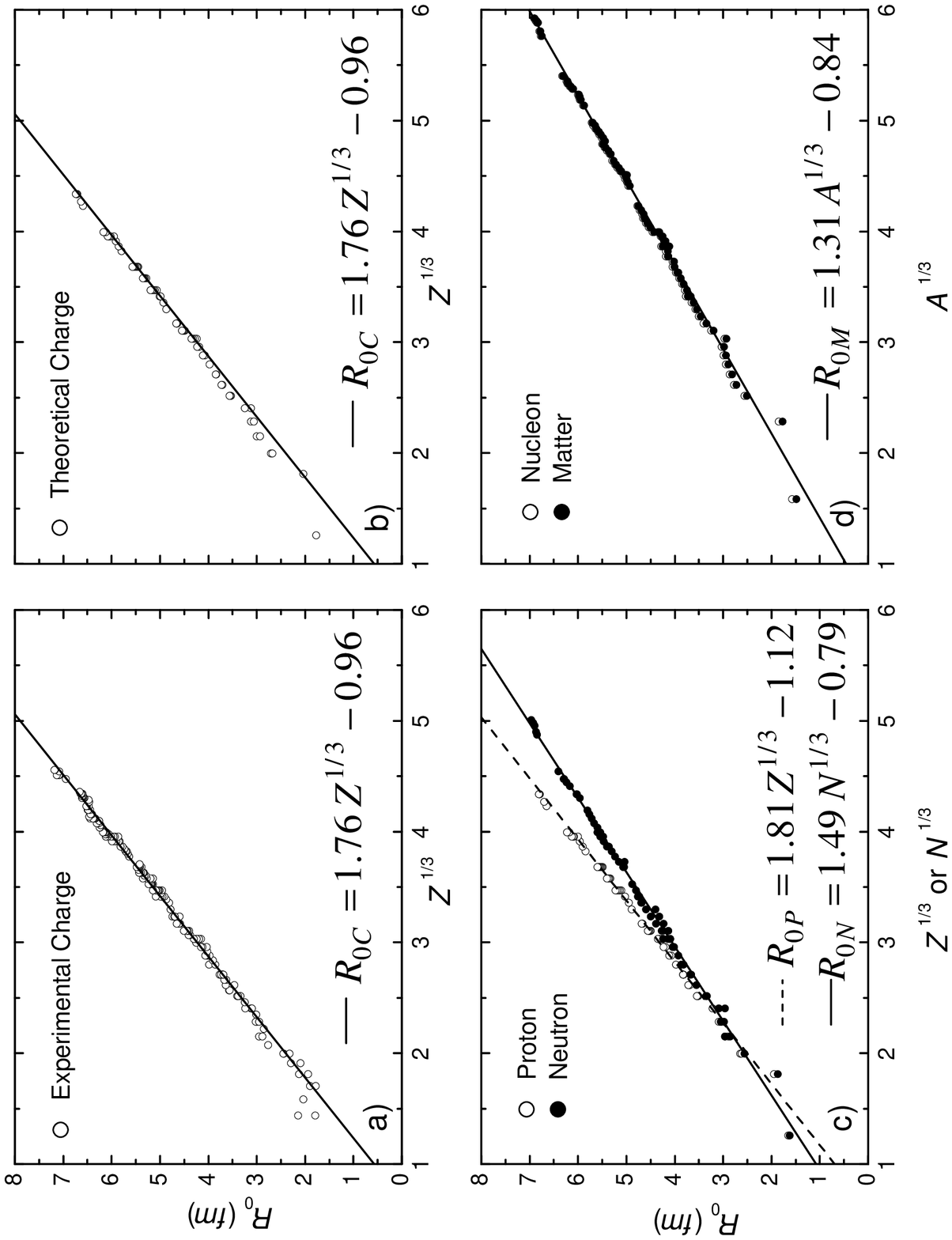}
\vspace{11.0cm}
\noindent
\caption{The $R_0$ parameter obtained for charge distributions extracted from 
electron scattering experiments and for theoretical densities obtained from 
Dirac-Hartree-Bogoliubov calculations.}
\end{figure}

\newpage

\begin{figure} 
\vspace{0.0cm}
\hspace{2.0cm}
\includegraphics{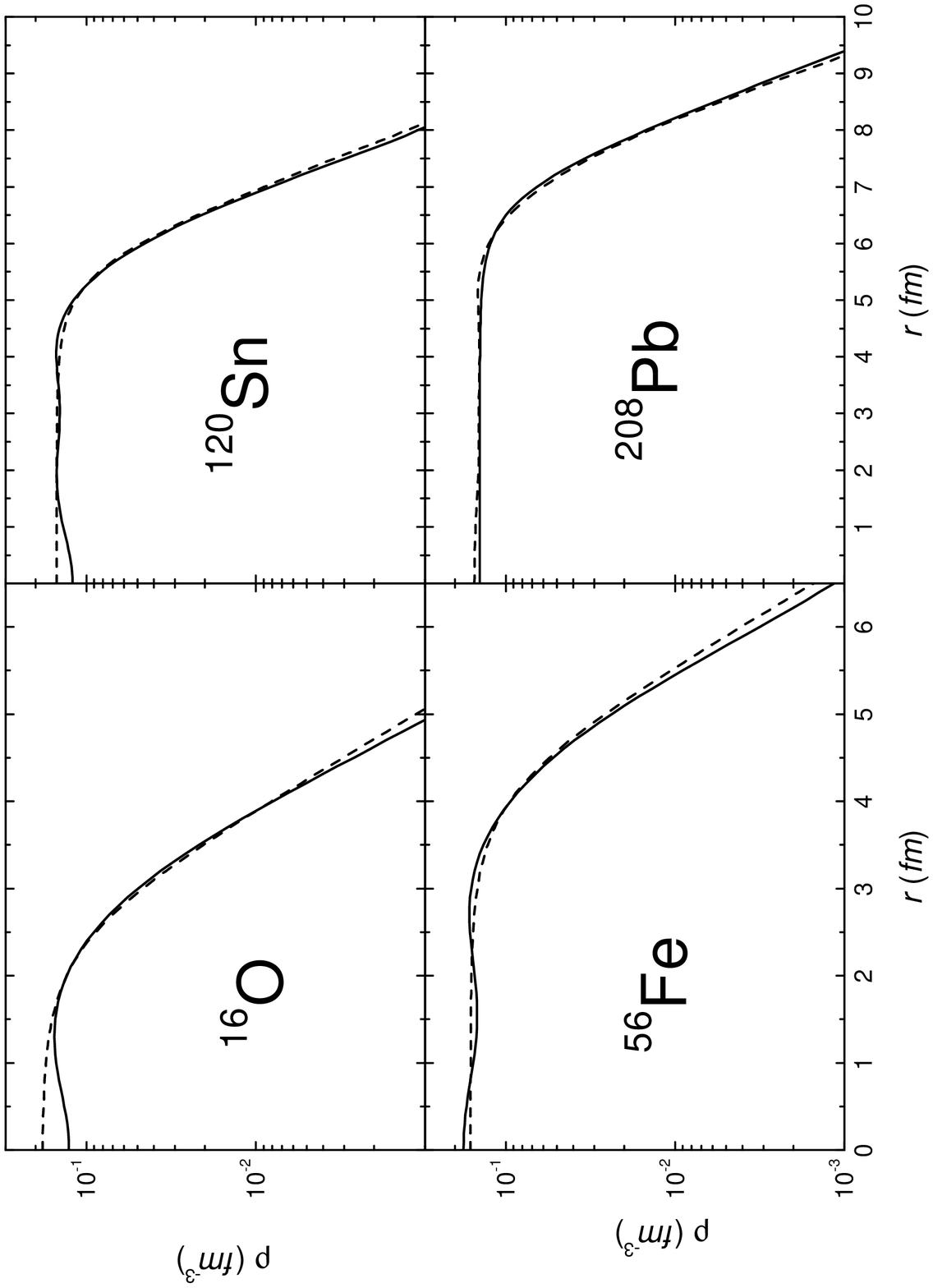}
\vspace{11.0cm}
\noindent
\caption{Nucleon densities from Dirac-Hartree-Bogoliubov calculations (solid 
lines) compared with the corresponding two-parameter Fermi distributions 
(dashed lines), with $a=0.50 \; fm$ and $R_0$ obtained through Eq. 6.}
\end{figure}

\newpage

\begin{figure} 
\vspace{0.0cm}
\hspace{2.0cm}
\includegraphics{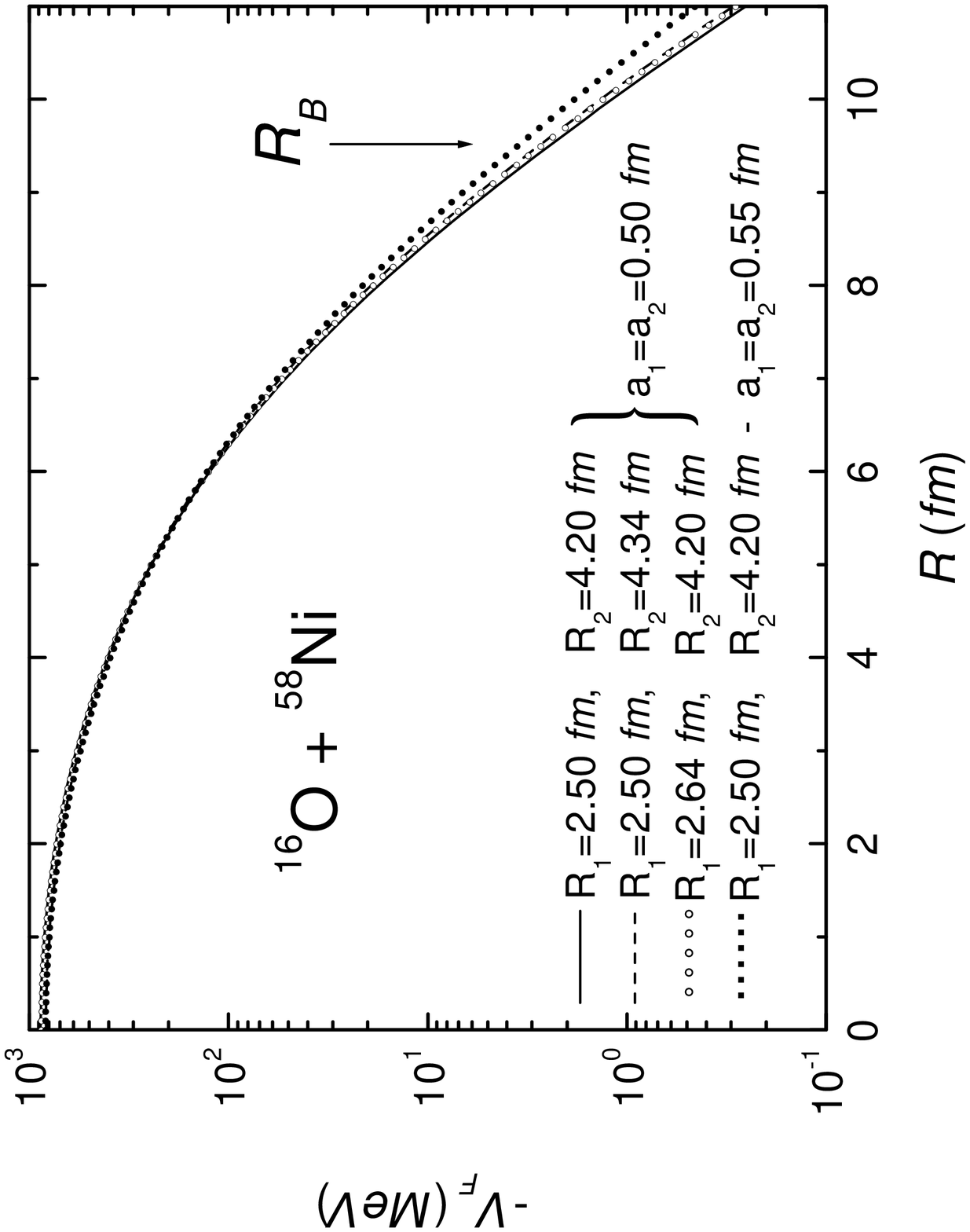}
\vspace{11.0cm}
\noindent
\caption{Folding potential for different sets of 2pF densities that may 
represent the $^{16}$O + $^{58}$Ni system. The approximate position of the 
s-wave barrier radius ($R_B$) is indicated in the figure.}
\end{figure}

\newpage

\begin{figure} 
\vspace{0.0cm}
\hspace{2.0cm}
\includegraphics{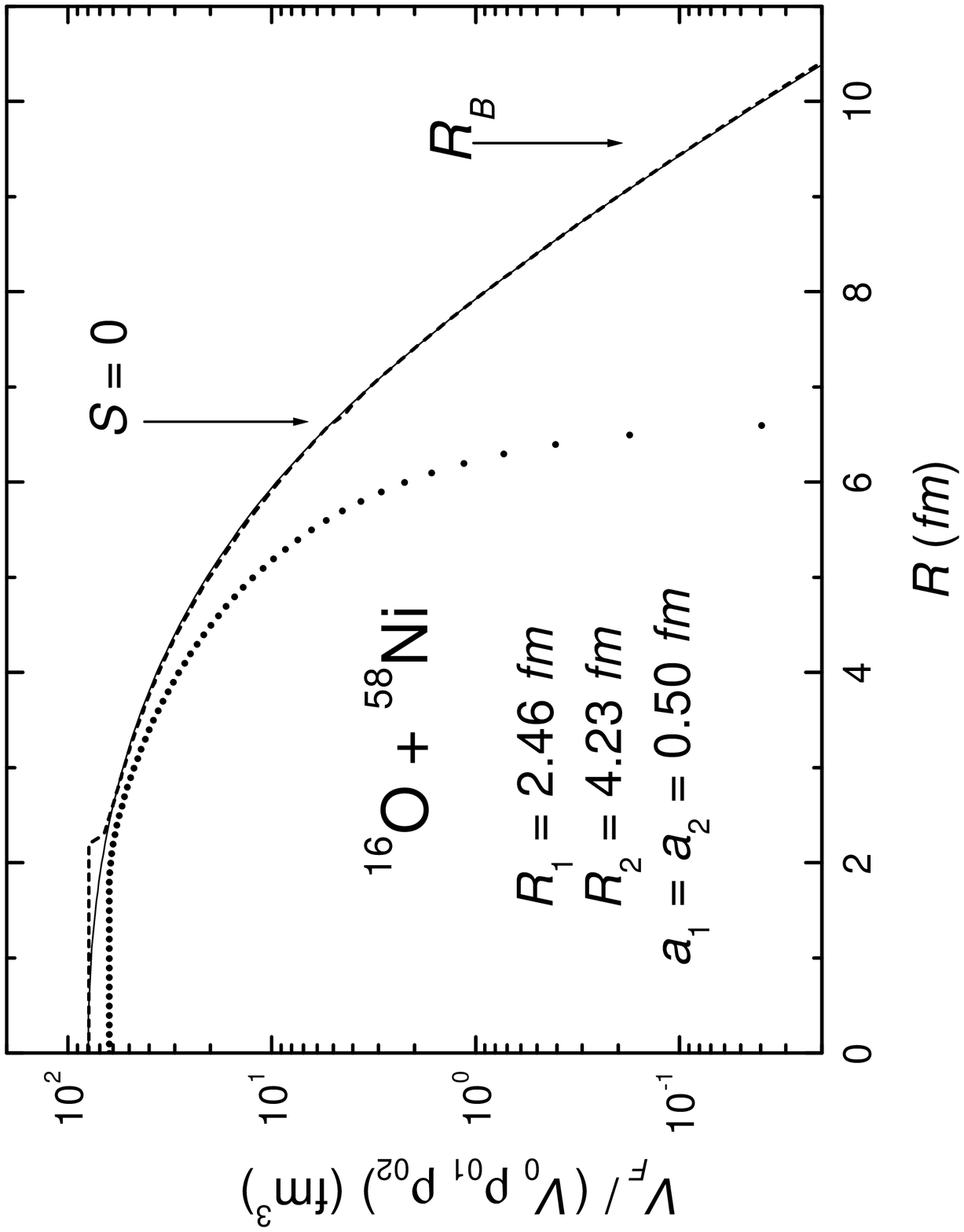}
\vspace{11.0cm}
\noindent
\caption{Folding potential in the zero-range approach calculated from 
numerical integration of Eq. 8 (solid line), for 2pF densities that may 
represent the $^{16}$O + $^{58}$Ni system. The dashed line represents the 
approximate analytical expressions, Eqs. 15, 16 and 17, while the dotted line 
concerns the exact result for $a=0$, Eqs. 9, 10 and 11. The approximate 
position of the s-wave barrier radius ($R_B$) and of the distance $R=R_1+R_2$ 
($s=0$) are indicated in the figure.}
\end{figure}

\newpage

\begin{figure} 
\vspace{0.0cm}
\hspace{2.0cm}
\includegraphics{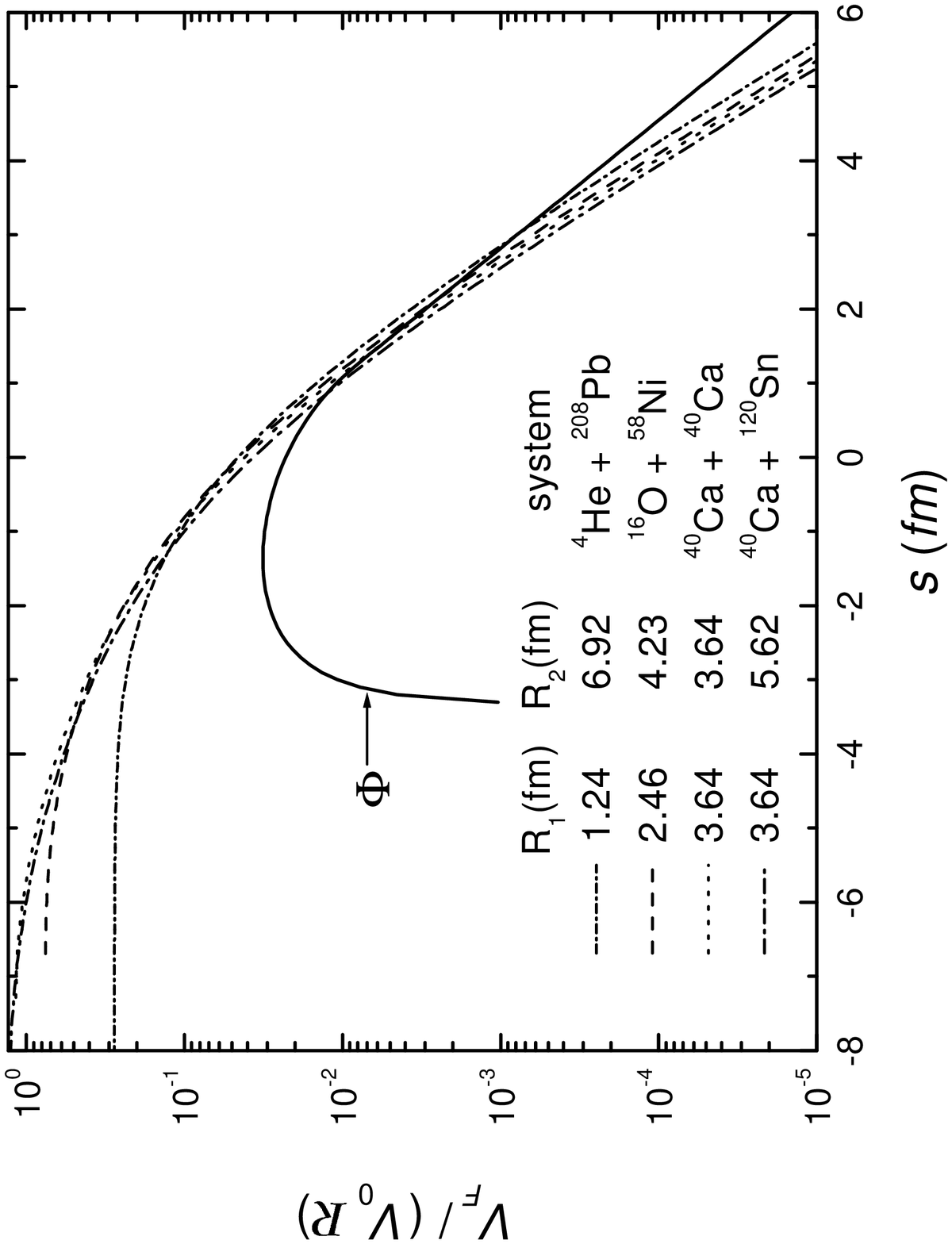}
\vspace{11.0cm}
\noindent
\caption{Normalized folding potential $V_F/(V_0 {\cal R})$ in the zero-range 
approach (Eq. 8) as a function of the distance $s=R-(R_1+R_2$), for several 
sets of 2pF distributions (with $a=0.50 \; fm$) that may represent the systems 
indicated in the figure. The proximity universal function $\Phi$ is also 
presented in arbitrary units.}
\end{figure}

\newpage

\begin{figure} 
\vspace{0.0cm}
\hspace{2.0cm}
\includegraphics{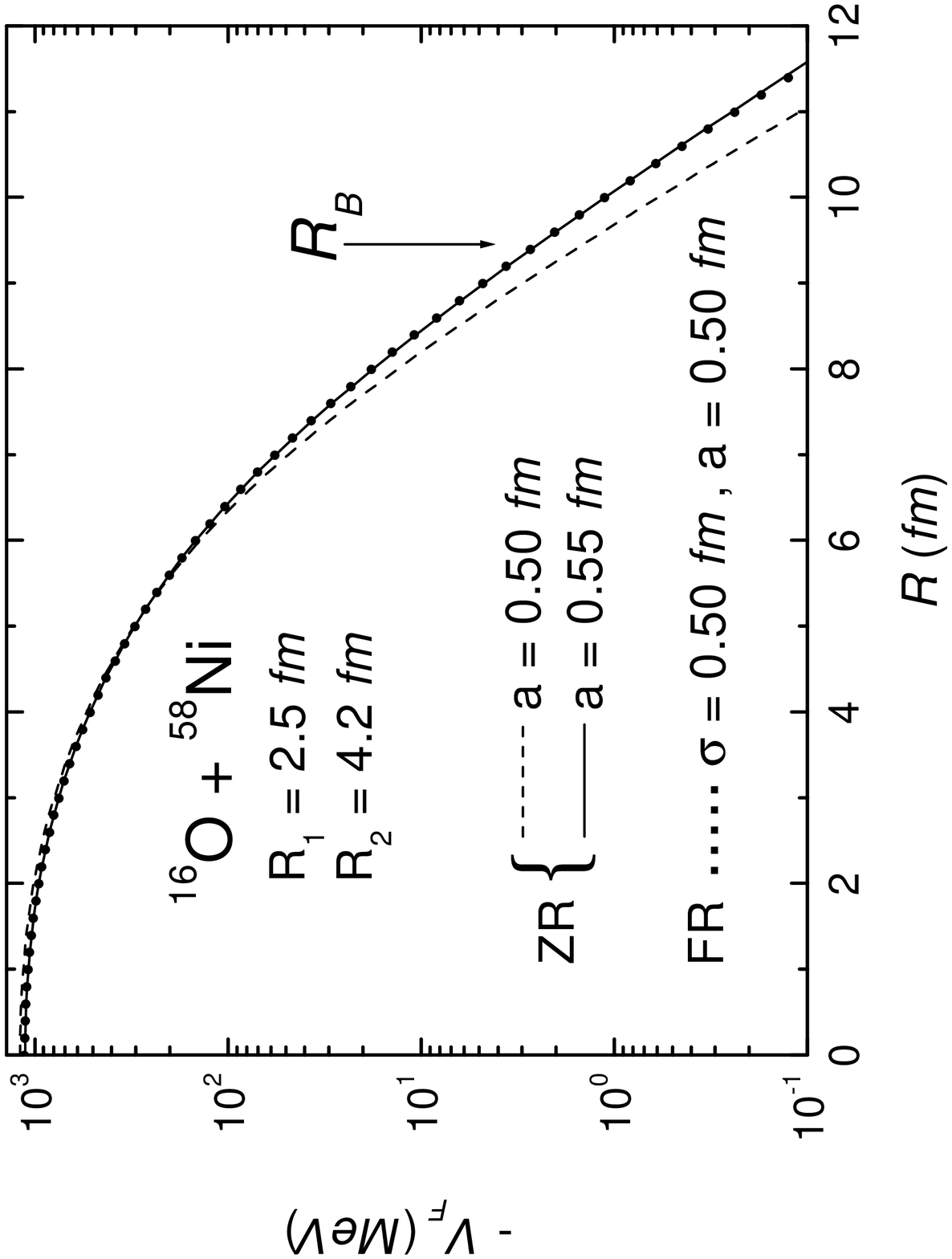}
\vspace{11.0cm}
\noindent
\caption{Double-folding potentials for 2pF distributions with different 
diffuseness values ($a$) that may represent the $^{16}$O + $^{58}$Ni system. 
The potentials have been calculated in the zero-range approach (ZR) or with a 
finite-range (FR) Yukawa function for the effective nucleon-nucleon 
interaction.}
\end{figure}

\newpage

\begin{figure} 
\vspace{0.0cm}
\hspace{2.0cm}
\includegraphics{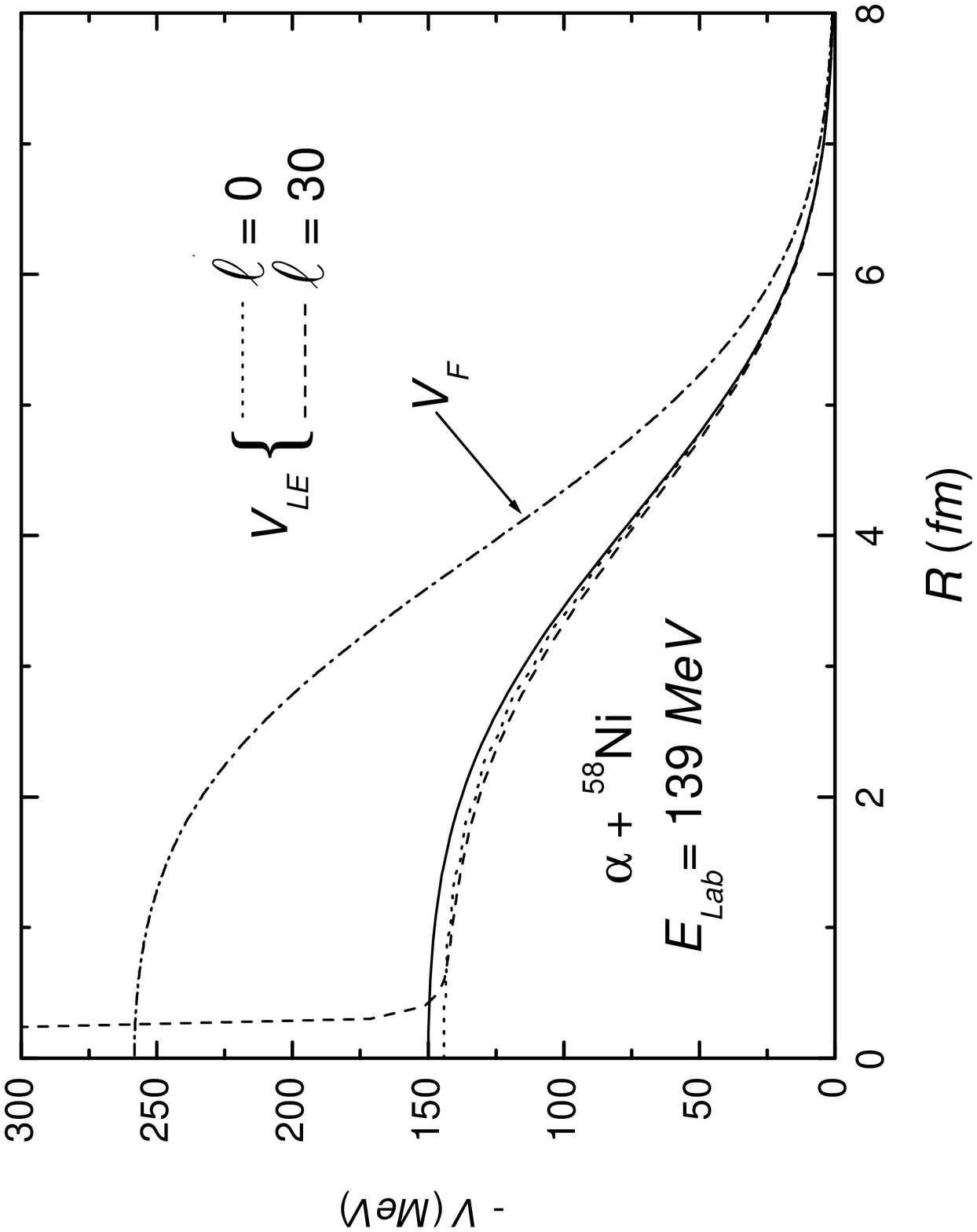}
\vspace{11.0cm}
\noindent
\caption{Double-folding ($V_F$) and $\ell$-dependent local-equivalent 
($V_{LE}$) potentials for the $\alpha$ + $^{58}$Ni system at 
$E_{Lab}=139 \; MeV$. The solid line represents the approximate expression, 
Eq. 34, for the local-equivalent potential.}
\end{figure}

\newpage

\begin{figure} 
\vspace*{15.0cm}
\hspace*{2.5cm}
\includegraphics{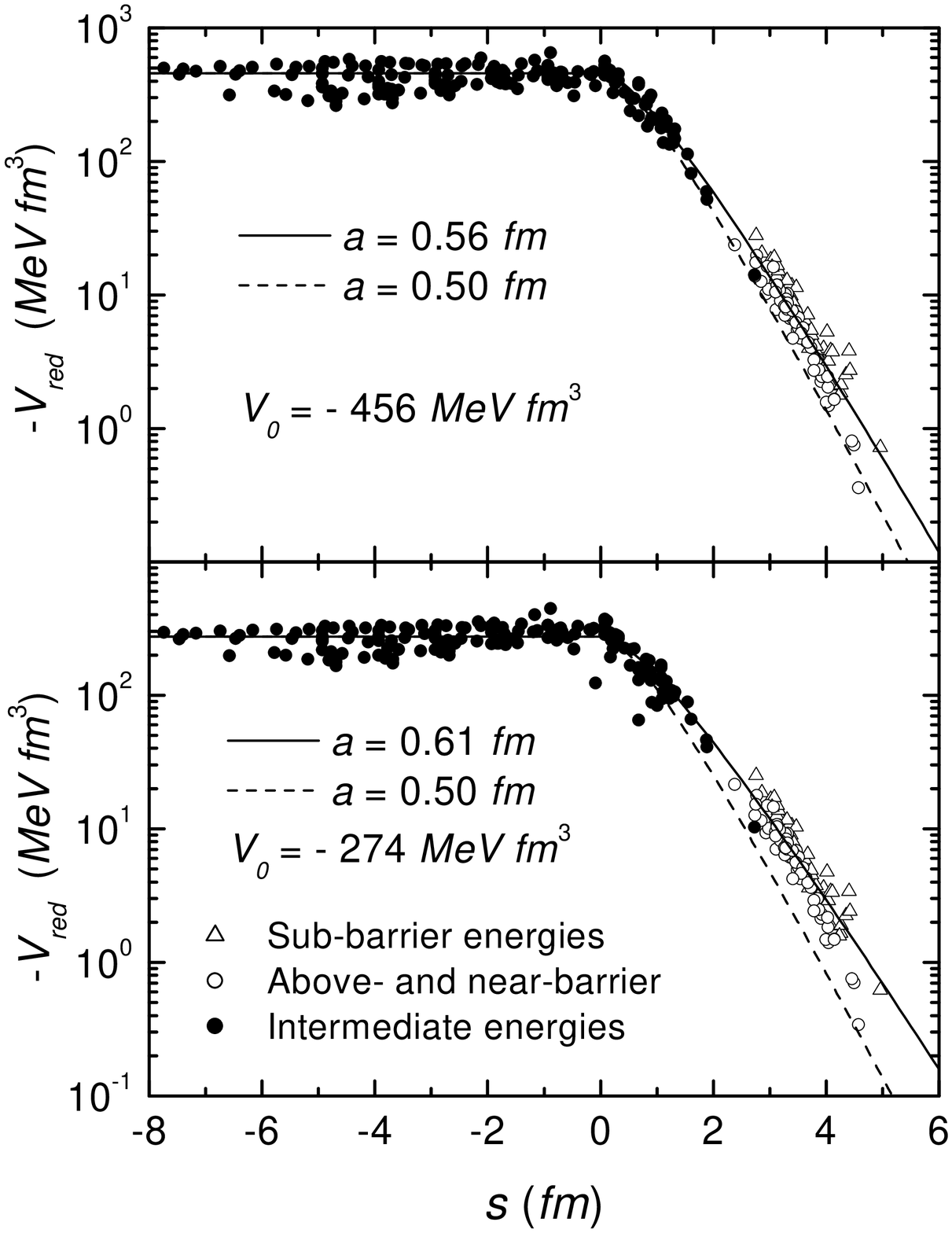}
\vspace{1.0cm}
\noindent
\caption{Experimental and theoretical reduced potentials in the context of the 
zero-range approach, with (top) or without (bottom) considering in the 
calculations the energy-dependence of the local-equivalent potential (Eq. 34) 
that arises from the Pauli nonlocality.}
\end{figure}

\newpage

\begin{figure} 
\vspace*{15.0cm}
\hspace*{2.5cm}
\includegraphics{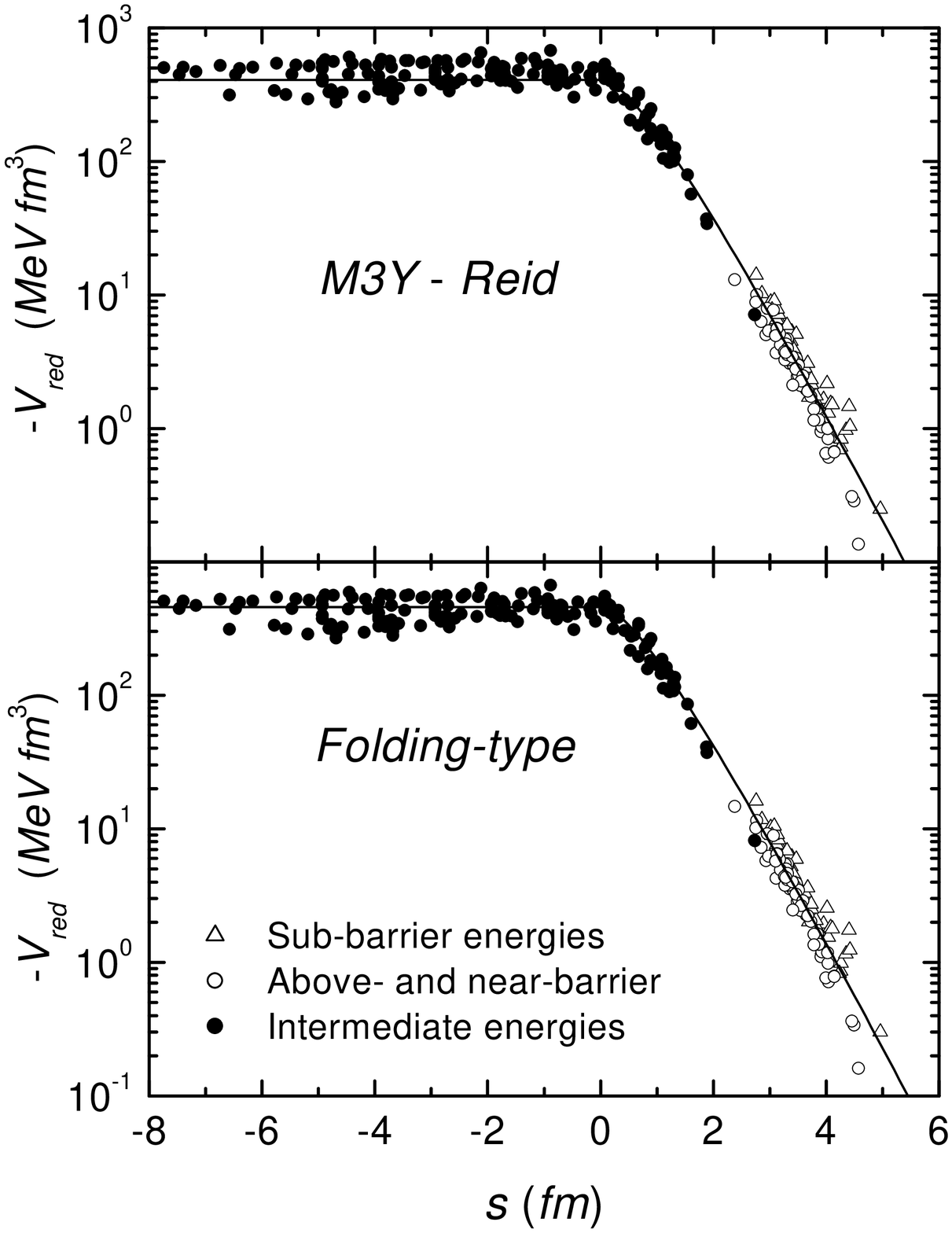}
\vspace{1.0cm}
\noindent
\caption{Comparison between experimental and theoretical reduced potentials in 
the context of the finite-range approach, with a M3Y-Reid (top) or folding-type 
(bottom) effective nucleon-nucleon interaction.}
\end{figure}

\newpage

\begin{figure} 
\vspace*{0.0cm}
\hspace*{2.0cm}
\includegraphics{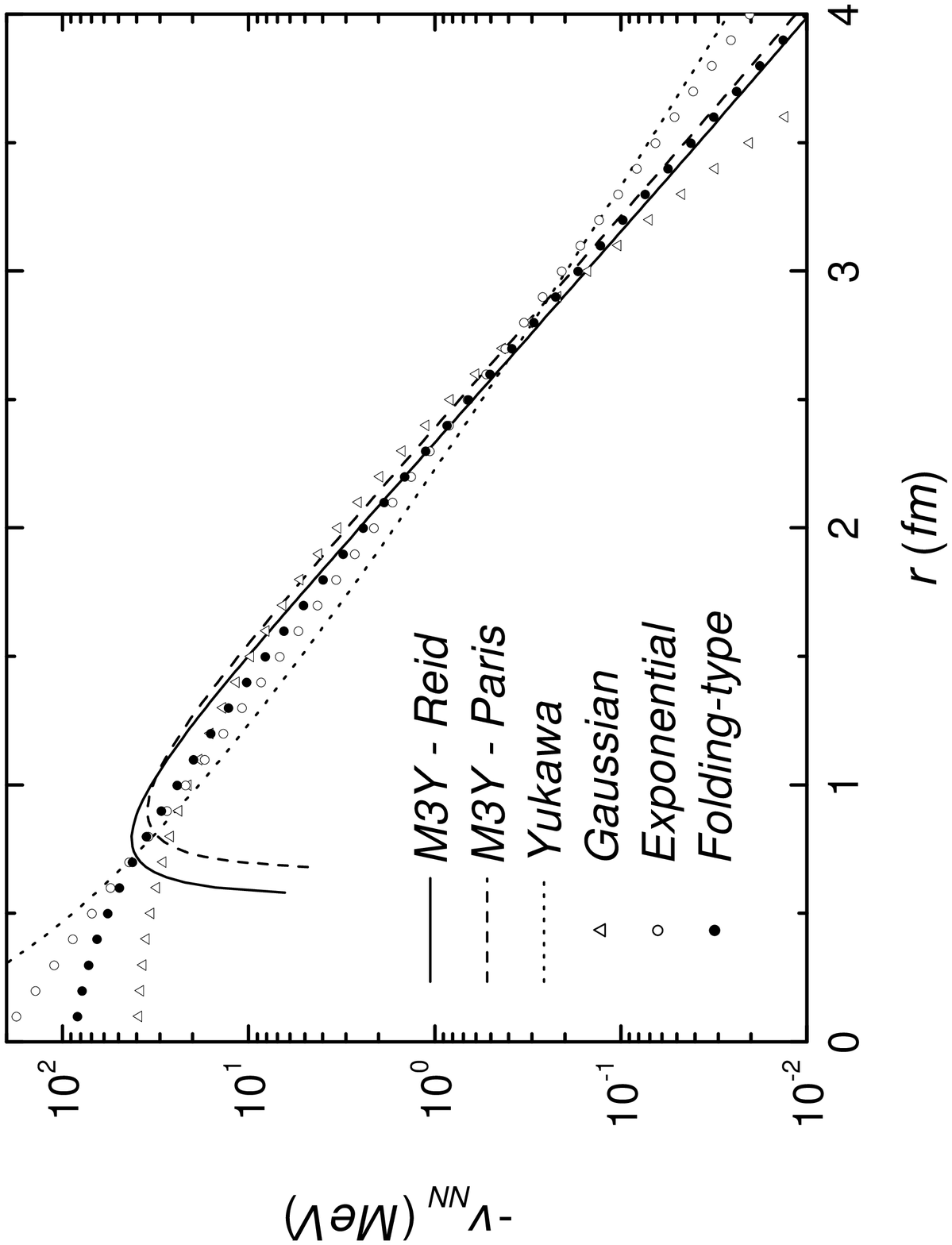}
\vspace{11.0cm}
\noindent
\caption{The complete set of effective nucleon-nucleon interactions considered 
in this work.}
\end{figure}

\end{document}